\documentclass[preprint,preprintnumbers,amsmath,amssymb]{revtex4}
\usepackage{amsfonts}    
\usepackage{amssymb}
\usepackage{latexsym}
\usepackage{eepic}
\usepackage{graphicx}
\usepackage[usenames,dvipsnames]{color}

\usepackage[active]{srcltx}

\newcommand{\la}{\lambda}

\newcommand{\De}{\Delta}

\newcommand{\rar}{\rightarrow}

\begin{document}

\title{On $1/Z$ expansion for two-electron systems}

\author{J.C.~Lopez Vieyra}
\email{vieyra@nucleares.unam.mx}
\author{A.V.~Turbiner}
\email{turbiner@nucleares.unam.mx}
\affiliation{Instituto de Ciencias Nucleares, Universidad Nacional
Aut\'onoma de M\'exico, Apartado Postal 70-543, 04510 M\'exico,
D.F., Mexico}

\begin{abstract}
The $1/Z$-expansion for the Coulomb system of infinitely massive center of charge Z and two electrons is discussed. Numerical deficiency in Baker et al, {\em Phys. Rev. \bf A41}, 1247 (1990) is indicated which continue to raise doubts in correctness of their calculations of the higher order coefficients in $1/Z$-expansion expressed in Refs.[4-5]. It is shown that a minor modification of a few first coefficients found in Ref.[3] allows to calculate the ground state energies at $Z\ =\ 1, 2,\ldots \ 10$ (as well as at $Z > 10$) with a portion of 15th decimal digit in comparison with highly accurate calculations by C.~Schwartz and by Nakashima-Nakatsuji. Ground state energies of two-electron ions $Z=11\ (Na^{9+})$ and  $Z=12\ (Mg^{10+})$ are found with 14 decimal digits.

\vskip 2cm

\centerline{\large arXiv:1309.2707 (quant-ph)}

\end{abstract}

\pacs{31.15.Pf,31.10.+z,32.60.+i,97.10.Ld}

\maketitle

Two electron system with infinitely-massive charge center $Z$ is described by the Hamiltonian
\begin{equation}
\label{H}
    {\cal H}\ =\ -\frac{1}{2} (\De_1 + \De_2) \ -\ \frac{Z}{r_{1}}\ -\
    \frac{Z}{r_{2}}  \ +\ \frac{1}{r_{ab}}\ .
\end{equation}
A change of variables in (\ref{H}), $\vec{r} \rar \vec{r}/Z$,\ leads to a new form of the Hamiltonian
\begin{equation}
\label{H_t}
    {\cal H}_t\ =\ -\frac{1}{2} (\De_1 + \De_2)\ -\ \frac{1}{r_{1}}\ -\
    \frac{1}{r_{2}}  \ +\  \frac{\la}{r_{ab}}\ , \quad \la = \frac{1}{Z}\ ,
\end{equation}
where the new energy $\tilde E(\la)=\frac{E(Z)}{Z^2}$. One of the important tools to study the spectra of (\ref{H_t}) is to develop the perturbation theory in powers of $\frac{1}{Z}$ constructing the expansion of $\tilde E$,
\begin{equation}
\label{E_t}
    \tilde E\ =\ \sum_{n=0}^{\infty} e_n \la^n\ .
\end{equation}
The first two coefficients are found analytically, $E_0 = -1, E_1 = 5/8$ while other coefficients can be found only approximately. The first attempt to calculate the next three of them was carried out by Hylleraas \cite{Hylleraas:1930} and then many workers dedicated a plenty of efforts to find as many of these coefficients as possible with the highest possible accuracy (see \cite{Silverman:1981} and \cite{Baker:1990} where extensive discussion with extended bibliography together with historical account are presented). A culmination of this story had happened at 1990 when Baker et al, \cite{Baker:1990} computed as many as $\sim 401$ coefficients of $1/Z$-expansion (\ref{E_t}) essentially overpassing all previous calculations in both accuracy and the number of coefficients.

It is worth mentioning one of the main reasons of such an interest to $1/Z$-expansion: it is among very few convergent(!) expansions in quantum physics, thus, it was considered as a challenge to find the radius of convergency $\la_*$. Furthermore, it was conjectured that the radius of convergency
coincides with inverse critical charge $\la_*=1/Z_{cr}$, for which the system $(Z, 2e)$ at $Z < Z_{cr}$ gets unbound (see \cite{Baker:1990} and references therein).

We are not aware about any studies of $1/Z$-expansion performed, after the paper of Baker et al, \cite{Baker:1990} was published, for about twenty years. In 2010 the results of \cite{Baker:1990} were challenged \cite{Zamastil:2010}. It was shown that the asymptotic behavior of the coefficients $e_n$ at $n$ tends to infinity derived from the analysis of the coefficients $e_n$, taken from $n=13$ to $n=19$, differs from one obtained from the analysis of the coefficients $e_n$ taken from $n=25$ to $n=401$. It leads to a significant deviation at large $n$ coefficients, e.g. at $n=200$ the discrepancy in the leading significant digit is about 50$\%$ while $e_{200}$ is of the order $10^{-16}$. In \cite{TG:2011} it was considered as an indication that the computational accuracy at \cite{Baker:1990} is exaggerated, in particular, the proclaimed quadruple precision arithmetics did not really work { (or insufficient)}, at least, for calculation of { the first significant digits} in the higher order coefficients. In \cite{TG:2011} it was also constructed the Puiseux expansion (in fractional degrees) of the ground state energy near the critical charge $Z_{cr}$. It was shown that the asymptotic behavior of the $e_n$ coefficients found in \cite{Baker:1990} is incompatible (slightly) with one derived from the Puiseux expansion, hence, the { Reinhardt} conjecture $\la_*=1/Z_{cr}$ fails.

The goal of this short Note is to check compatibility of the ground state energies at
$Z=1, 2, 3, \ldots , 10$ found perturbatively, in particular, using the coefficients
$e_n$ obtained { (and published)} in \cite{Baker:1990} with highly accurate results for the ground state energies of two-electron ions obtained in \cite{Schwartz:2006} and \cite{Nakashima:2007}. All calculations were made and checked in three different multiple precision arithmetics: (i) ifort q-precision real*16 (quadruple precision), (ii) Maple Digits=30 in Maple 13 and (iii) Charles Schwartz multiple precision arithmetic package.

The first observation is that we do not confirm the statement
from Baker et al, \cite{Baker:1990} (p.1254):

\noindent
\textit{The sum of the $e_n$'s for $n$ running from 0 to 401 is
\[
 -0.527 \, 751 \, 016 \, 544 \, 266
\]
which at the time we did our calculations was the most accurate estimate of the energy for the ground state of H$^-$}. Our result { using $e_n$ published in \cite{Baker:1990}}
\[
 -0.527 \, 751 \, 016 \, 544 \, 160
\]
differs in the last three decimal digits. It gives us a chance to suspect that the quadruple precision arithmetics used by the authors of \cite{Baker:1990} did not really work and, probably, was downgraded to double precision arithmetics, { or was insufficient to go beyond} 12 decimal digits. Thus, we can not trust decimal digits in coefficients $e_n$ beyond 12th digit and, in particular, all $e_n$ for $n > 135$ (when the (rounded) coefficients are of the order -12 and less, see \cite{Baker:1990}, Table III) are unreliable. Probably, they are random numbers ranging from 0 to 9. Concluding, we note that both above numbers coincide up to 12 decimal digits with accurate result given by \cite{Nakashima:2007} for $Z=1$
\[
  -0.527 \, 751 \, 016 \, 544 \, 377
\]
but differ from it in 13th { and subsequent decimal digits. It is the explicit indication that $e_n$ calculated in \cite{Baker:1990} beyond 12 decimal digits are wrong}.

Taking all 401 $e_n$ coefficients from \cite{Baker:1990} we calculated the weighted sums  (\ref{E_t}) for $Z=1, \ldots, 10$ and compared with the energies found in \cite{Schwartz:2006} at $Z=2$ and \cite{Nakashima:2007}. They coincided up to 12 decimal digits. We consider it as an indication that 12 decimal digits in coefficients $e_n$ in \cite{Baker:1990}, Table III might be correct. Taking $e_2$ from \cite{Schwartz:2013} and making a "minimal" modification of the coefficients $e_3, e_4, e_5, e_6$, see Table~\ref{bakermodified}, we obtain a substantial improvement in the agreement between energies found perturbatively through (3) and ones from \cite{Schwartz:2006} and \cite{Nakashima:2007}: for all $Z=1 \ldots 10$ they differ in a portion of 15th decimal digit (see Table \ref{E}). Since a contribution of the higher order coefficients in the weighted sum (\ref{E_t}) to the ground state energy decreases dramatically with the increase of the charge $Z$, it is guaranteed that the same number 14 of the correct decimal digits should be obtained for larger $Z > 10$. As an example, the ground state energies for two-electron ions $Z=11\ (Na^{9+})$ and  $Z=12\ (Mg^{10+})$  are calculated perturbatively and presented in Table \ref{E}. They essentially improve the most accurate results, known up to date to the present authors, by \cite{Thakkar:1977}, they differ from them in the sixth decimal digit.
\begin{table}[htb]
\begin{center}
 \begin{tabular}{ll}\hline
$e_0 =$ & $-$1\\
$e_1 =$ & $+$5/8\\
$e_2 =$ & $-$0.157\ 666\ 429\ 469\ 1{\bf 5} ${}^{s}$  \\
$e_3 =  $ & $+$0.008\ 699\ 031\ 52{\bf 7}\ {\bf 90} \ (8 - -)   \\
$e_4 =  $ & $-$0.000\ 888\ 707\ 284\ 2{\bf 3} \ (-)  \\
$e_5 =  $ & $-$0.001\ 036\ 371\ 848\  {\bf 05} \ (- -)  \\
$e_6 =  $ & $-$0.000\ 612\ 940\ 520\ 5{\bf 3} \ (-) \\
$e_7 =  $ & $-$0.000\ 372\ 175\ 576\ 5 \\
$e_8 =  $ & $-$0.000\ 242\ 877\ 973\ 2 \\
$e_9 =  $ & $-$0.000\ 165\ 661\ 054\ 7 \\
$e_{10} =    $ & $-$0.000\ 116\ 179\ 202\ 6 \\
\hline
\end{tabular}
\caption{\label{bakermodified} List of the first perturbation coefficients $e_n$, some of them are modified in comparison with ones found in \cite{Baker:1990}, modification is marked by bold, non-modified (original) digits from \cite{Baker:1990} are shown in brackets, dash means that digit was not present in original coefficient; { ${}^{s}$ $e_2$ found by C Schwartz \cite{Schwartz:2013}: the difference with $e_2$ from \cite{Baker:1990} in 14th decimal digit, 4 is replaced by 5.}}
\end{center}
\end{table}
\begin{table}[hbt]
\begin{center}
{\small
\begin{tabular}
{l|rr}
\hline
$Z$ & $E$ (a.u.) from (\ref{E_t}) \quad &\quad $E$ (a.u.) from Refs.\cite{Schwartz:2006},  \cite{Nakashima:2007}\\
\hline
1  &  -0.527 751 016 544 380  &  -0.527 751 016 544 377          \\
2  &  -2.903 724 377 034 122  & -2.903 724 377 034 119           \\
3  &  -7.279 913 412 669 304  & -7.279 913 412 669 305           \\
4  & -13.655 566 238 423 584  & -13.655 566 238 423 586          \\
5  & -22.030 971 580 242 778  & -22.030 971 580 242 781          \\
6  & -32.406 246 601 898 527  & -32.406 246 601 898 530          \\
7  & -44.781 445 148 772 701  & -44.781 445 148 772 704          \\
8  & -59.156 595 122 757 921  & -59.156 595 122 757 925          \\
9  & -75.531 712 363 959 486  & -75.531 712 363 959 491          \\
10 & -93.906 806 515 037 544  & -93.906 806 515 037 549          \\
11 & -114.281 883 776 072 717 &      -114.281 879\  $(*)$           \\
12 & -136.656 948 312 646 925 &      -136.656 944\  $(*)$           \\
 \hline
\end{tabular}
}
\end{center}
\caption{\label{E} Left column: perturbatively found energies $E(Z)$ from (\ref{E_t}) with coefficients $e_{3,4,5,6}$, see Table~\ref{bakermodified} and others from \cite{Baker:1990}.
Right column: rounded results from Refs. \cite{Schwartz:2006} $(Z=2)$ and \cite{Nakashima:2007}, the results from Ref. \cite{Thakkar:1977} marked by $(*)$}
\end{table}

Concluding we want to state that independent calculation of coefficients of $1/Z$-expansion is needed. From one side, it can be repeated the same calculation as in \cite{Baker:1990} but with reliable multiple precision package. From another side, much more accurate trial functions, than one used in \cite{Baker:1990}, are now available like one by Korobov \cite{Korobov:2000}, or by Drake et al, \cite{Drake:2002}, or by Schwartz \cite{Schwartz:2006} (see {\it Notes added}), or by Nakashima-Nakatsuji \cite{Nakashima:2007} among others. It seems any of these trial functions can be taken as entry in the procedure elaborated in \cite{Baker:1990}. Such a calculation can eventually allows to find the asymptotic behavior $e_n$ coefficients, radius of convergency of $1/Z$-expansion (a value of critical charge) and reveal a structure of singularity. { Besides that we feel a need to develop analytical approach for finding the asymptotic behavior of coefficients, probably, similar to dispersion relations in a coupling constant for anharmonic oscillators due to Bender and Wu \cite{BW}}.

\textit{\small Acknowledgements}.  Authors want to express their gratitude to C.~Schwartz for
providing multiple precision package, useful discussions and providing results before publication \cite{Schwartz:2013,Schwartz:2013n}. The research is supported in part by DGAPA grant IN109512 (Mexico). Authors thanks the University Program FENOMEC (UNAM, Mexico) for partial support.

\noindent
{
\textit{Note added.I.\ (Sept.22, 2013)}.\ When the article was submitted for publication, C.~Schwartz informed us that he carried out a direct calculation of $e_2$ with 25 decimal digits \cite{Schwartz:2013}
\[
e_2 = -0.15766 64294 69150 94105 66793 \quad .
\]
It does not confirm the 14th decimal digit found in \cite{Baker:1990}. Thus, it provides one more indication that the accuracy reported in \cite{Baker:1990} was exaggerated and the results presented there are not reliable. We used the $e_2$ reported in \cite{Schwartz:2013} for our analysis, see Table I.

\textit{Note added.II.\ (Nov.21, 2013)}.\ When the article was under consideration, C.~Schwartz informed us that he carried out a direct calculation of the coefficients $e_3 - e_{20}, e_{30}, e_{40}, e_{50}$ \cite{Schwartz:2013n}. It was used the $F$-function method \cite{Schwartz:2006} with basis set length up to 3091 terms
\footnote{For comparison, the basis length in \cite{Baker:1990} was 476 terms}
in 58-digit arithmetics. It was shown that the 12th decimal digit (and next ones when available) of $e_3 - e_{20}$ calculated in \cite{Baker:1990} are found systematically wrong which is in agreement with main conclusion of this paper. Furthermore, the number of correct significant digits reduces gradually with $n$ and becomes for $e_{50}$ equal to 5 out of 12 decimal digits.
If such a tendency will continue there will be no correct significant digits found in $e_n$ for $n \gtrsim 130$.
}


\begin{thebibliography}{99}

\bibitem{Hylleraas:1930}
         E.A.~Hylleraas,
         {\em Z.Phys. \bf 65}, (1930) 209

\bibitem{Silverman:1981}
         J.N.~Silverman,
         {\em Phys. Rev. \bf A23}, 441 (1981)

\bibitem{Baker:1990}
         J.D.~Baker, D.E.~Freund, R.N.~Hill, and J.D.~Morgan III,
         {\em Phys. Rev. \bf A41}, 1247 (1990)

\bibitem{Zamastil:2010}
         J.~Zamastil, J.~Cizek, L.~Skala, and M.~Simanek,
         {\em Phys. Rev. \bf A81} (2010) 032118

\bibitem{TG:2011}
         A.V.~Turbiner, and N.L.~Guevara,
        {\it Phys Rev \bf A84} (2011) 064501 (4pp)

\bibitem{Schwartz:2006}
        C.~Schwartz,
        {\it Int. J. Mod. Phys. \bf E15}, 877 (2006); \\
        e-Print arXiv: physics/0208004, math-ph/0605018

\bibitem{Nakashima:2007}
        H.~Nakashima, and H.~Nakatsuji,
        {\it J. Chem. Phys. \bf 127}, 224104 (2007)

\bibitem{Thakkar:1977}
         A.J.~Thakkar, and V.H.~Smith, Jr.
         {\em Phys. Rev. \bf A15}, 1-15 (1977)

\bibitem{Korobov:2000}
         V.I.~Korobov,
         {\em Phys. Rev. \bf A61} (2000) 064503; {\em ibid \bf A66} (2002) 024501

\bibitem{Drake:2002}
        G.~W.~F.~Drake, M.~M.~Cassar, and R.~A.~Nistor,
        {\em Phys. Rev. \bf A65}, 054501 (2002)

\bibitem{BW}
         C.M.~Bender, T.T.~Wu,
         \textit{Anharmonic Oscillator}, {\it Phys. Rev. \bf 184}, 1231 (1969);
         \textit{Anharmonic Oscillator.II}, {\it Phys. Rev.  \bf D 7} , 1620 (1973)

\bibitem{Schwartz:2013}
        C.~Schwartz, {\it Private communication} (Sept.22, 2013)

\bibitem{Schwartz:2013n}
        C.~Schwartz, {\it Private communication} (Nov.15, 2013)



\end{thebibliography}
\end{document}